\documentclass{wqcd03}                 % twocolumn proceedings style
\usepackage{txfonts}                   % if you have it, to get Times family
\usepackage{pennames}                  % copy  pennames.sty in working directory
\confname{QCD@Work 2003 - International Workshop on QCD, Conversano, Italy, 
14--18 June 2003}

\title{Transverse mass distributions of strange particles 
       produced in Pb-Pb collisions at 158 $A$\ GeV/$c$}

\author{G.~E.~Bruno\addressmark{b} for the NA57 Collaboration:\\
F~Antinori\addressmark{l},
P~Bacon\addressmark{e},
A~Badal{\`a}\addressmark{g},
R~Barbera\addressmark{g},
A~Belogianni\addressmark{a},
A~Bhasin\addressmark{e},
I~J~Bloodworth\addressmark{e},
M~Bombara\addressmark{j},
G~E~Bruno\addressmark{b},
S~A~Bull\addressmark{e},
R~Caliandro\addressmark{b},
M~Campbell\addressmark{h},
W~Carena\addressmark{h},
N~Carrer\addressmark{h},
R~F~Clarke\addressmark{e},
A~Dainese\addressmark{l},
A~P~de~Haas\addressmark{s},
P~C~de~Rijke\addressmark{s},
D~Di~Bari\addressmark{b},
S~Di~Liberto\addressmark{o},
R~Divia\addressmark{h},
D~Elia\addressmark{b},
D~Evans\addressmark{e},
K~Fanebust\addressmark{d},
G~A~Feofilov\addressmark{q},
R~A~Fini\addressmark{b},
P~Ganoti\addressmark{a},
B~Ghidini\addressmark{b},
G~Grella\addressmark{p},
H~Helstrup\addressmark{d},
A~K~Holme\addressmark{k},
A~Jacholkowski\addressmark{b},
G~T~Jones\addressmark{e},
P~Jovanovic\addressmark{e},
A~Jusko\addressmark{i},
R~Kamermans\addressmark{s},
J~B~Kinson\addressmark{e},
K~Knudson\addressmark{h},
A~A~Kolozhvari\addressmark{q},
V~Kondratiev\addressmark{q},
I~Kr\'alik\addressmark{i},
A~Krav\v c\'akov\'a\addressmark{j},
P~Kuijer\addressmark{s},
V~Lenti\addressmark{b},
R~Lietava\addressmark{f},
G~L\o vh\o iden\addressmark{k},
V~Manzari\addressmark{b},
G~Martinsk\'a\addressmark{j},
M~A~Mazzoni\addressmark{o},
F~Meddi\addressmark{o},
A~Michalon\addressmark{r},
M~Morando\addressmark{l},
F~Navach\addressmark{b},
P~I~Norman\addressmark{e},
A~Palmeri\addressmark{g},
G~S~Pappalardo\addressmark{g},
B~Pastir\v c\'ak\addressmark{i},
J~Pi\v s\'ut\addressmark{f},
N~Pisutova\addressmark{f},
E~Quercigh\addressmark{l},
F~Riggi\addressmark{g},
D~R\"ohrich\addressmark{c},
G~Romano\addressmark{p},
K~\v{S}afa\v{r}\'{i}k\addressmark{h},
L~\v S\'andor\addressmark{i},
E~Schillings\addressmark{s},
G~Segato\addressmark{l},
M~Sen\`e\addressmark{m},
R~Sen\`e\addressmark{m},
W~Snoeys\addressmark{h},
F~Soramel\addressmark{l},
M~Spyropoulou-Stassinaki\addressmark{a},
P~Staroba\addressmark{n},
T~A~Toulina\addressmark{q},
R~Turrisi\addressmark{l},
T~S~Tveter\addressmark{k},
J~Urb\'{a}n\addressmark{j},
F~F~Valiev\addressmark{q},
A~van~den~Brink\addressmark{s},
P~van~de~Ven\addressmark{s},
P~Vande~Vyvre\addressmark{h},
N~van~Eijndhoven\addressmark{s},
J~van~Hunen\addressmark{h},
A~Vascotto\addressmark{h},
T~Vik\addressmark{k},
O~Villalobos~Baillie\addressmark{e},
L~Vinogradov\addressmark{q},
T~Virgili\addressmark{p},
M~F~Votruba\addressmark{e},
J~Vrl\'{a}kov\'{a}\addressmark{j} and
P~Z\'{a}vada\addressmark{n}
} 
\address{
$^{a}$ Physics Department, University of Athens, Athens, Greece\\
$^{b}$ Dipartimento IA di Fisica dell'Universit{\`a}
e del Politecnico di Bari and INFN, Bari, Italy \\
$^{c}$ Fysisk Institutt, Universitetet i Bergen, Bergen, Norway\\
$^{d}$ H{\o}gskolen i Bergen, Bergen, Norway\\
$^{e}$ University of Birmingham, Birmingham, UK\\
$^{f}$ Comenius University, Bratislava, Slovakia\\
$^{g}$ University of Catania and INFN, Catania, Italy\\
$^{h}$ CERN, European Laboratory for Particle Physics, Geneva, Switzerland\\
$^{i}$ Institute of Experimental Physics, Slovak Academy of Science,
       Ko\v{s}ice, Slovakia\\
$^{j}$ P.J. \v{S}af\'{a}rik University, Ko\v{s}ice, Slovakia\\
$^{k}$ Fysisk Institutt, Universitetet i Oslo, Oslo, Norway\\
$^{l}$ University of Padua and INFN, Padua, Italy\\
$^{m}$ Coll\`ege de France, Paris, France\\
$^{n}$ Institute of Physics, Prague, Czech Republic\\
$^{o}$ University ``La Sapienza'' and INFN, Rome, Italy\\
$^{p}$ Dipartimento di Scienze Fisiche ``E.R. Caianiello''
       dell'Universit{\`a} and INFN, Salerno, Italy\\
$^{q}$ State University of St. Petersburg, St. Petersburg, Russia\\
$^{r}$ Institut de Recherches Subatomique, IN2P3/ULP, Strasbourg, France\\
$^{s}$ Utrecht University and NIKHEF, Utrecht, The Netherlands
}
\begin{document}
\begin{abstract}
Experiment NA57 has collected high statistics, high purity samples 
of \PKzS, \PgL, $\Xi$\ and  $\Omega$\ produced in Pb-Pb  
collisions at 158 $A$\ GeV/$c$. 
In this paper we present a %preliminary 
study of the transverse mass 
spectra of these particles for a  sample of events corresponding to 
about the most central 55\% of the inelastic Pb-Pb cross  section. 
We analyse the transverse mass distributions in the framework 
of the blast-wave model for the full sample under consideration 
and, for the first  time at the SPS, as a function of the event 
centrality.  
\end{abstract}
%
%% \maketitle needs to be after the author and address info and the abstract 
\maketitle
%
%% standard LaTeX from here on...
%
\section{Introduction}
The NA57 experiment has been designed to study the production of 
strange baryons and antibaryons in heavy ion collisions. The experiment 
has extended the WA97~\cite{WA97} measurements to a wider centrality 
range and to a lower beam momentum. 
NA57 results on hyperon yields~\cite{Ladislav} at 158 $A$\ GeV/$c$\ confirm 
the enhancement pattern observed by 
the WA97 experiment,  
%WA97,  
with an enhancement factor increasing with increasing 
strangeness.  
%content of the particle.
\newline
In this paper we shall concentrate on the analysis of the transverse mass
($m_{\tt T}=\sqrt{p_{\tt T}^2+m^2}$)
spectra for
the \PgL, \PgXm, \PgOm\ hyperons, their antihyperons and \PKzS\
measured in Pb-Pb collisions at 158 $A$\ GeV/$c$.
%The $m_{\tt T}$\
%spectra are expected to be sensitive to the details of the production
%dynamics~\cite{RafBook}.  
%
\section{The NA57 experiment}
The layout of the NA57 experiment has been described in detail 
elsewhere~\cite{NA57exp}. A telescope of compactly packed silicon 
pixel detectors is used as main tracking device. Additional  
pixel and double-sided silicon microstrip detectors 
are used to improve the momentum resolution for high momentum tracks.  
The telescope inclination angle with respect to the beam line 
and the distance %of the first pixel plane 
from the target are set to accept particles produced 
in about half a unit of rapidity
around central rapidity and medium 
%$p_T$. 
transverse momentum.  
\newline
An array of scintillation counters  
provides a fast signal to trigger on the centrality of the collisions.
A more precise centrality measurement is provided by two stations of 
silicon strip detectors.  All detectors are located  
in the 1.4 T field of the Goliath magnet. The triggered fraction of 
the total inelastic cross section is about 55\%.  
\newline
The \PKzS\ mesons, the \PgL, \PgXm\ and \PgOm\ hyperons and their antiparticles
are identified by reconstructing their weak
decays into final states containing only charged particles. Decays are 
required to take place in a fiducial volume located between a fixed distance 
from the target and the first plane of the telescope. The selection 
procedure allows for extraction of 
hyperon and \PKzS\ 
%particle  
signals with negligible 
background~\cite{tre}.   
\newline
%The measured charged particle multiplicity distribution  
%%, which is shown in figure~\ref{fig:multiplicity}, 
Events have been divided into  
% has been divided into 
five centrality classes (0,1,2,3,4), class 0  
being the most peripheral and class 4 the most central, 
according to the charge particle multiplicity measured in the  
pseudorapidity intervals $2<\eta<3$\ and $3<\eta<4$. 
%%The drop at very low multiplicities is due to the centrality 
%%selection applied at the trigger level.
The centrality of the collision is expressed as number of wounded nucleons
computed from the measured trigger cross sections using the Glauber model.
The centrality determination is described in detail  
in ref.~\cite{Multiplicity}.  
\section{Data sets and analysis}
The results presented in this paper are based on the analysis of a data
sample consisting of 460 M events of Pb-Pb collisions collected in the years
1998 and 2000. \newline
For each particle species we define the fiducial
acceptance window using a Monte Carlo simulation of the apparatus,
in order to exclude the border regions.  
%borders where the systematic errors are more difficult to evaluate.
\newline
All data are corrected for geometrical acceptance and for detector and 
reconstruction inefficiencies on a particle-by-particle basis, using 
a Monte Carlo technique where simulated particles are embedded into real 
events (the procedure has been described in detail in ~\cite{MonteCarlo}).  
%with a procedure that has been described in detail elsewhere~\cite{MonteCarlo}.  
\newline
The correction method %described above 
is CPU intensive; therefore, while all the reconstructed $\Xi$s and $\Omega$s 
have been individually weighted,  
for the much more abundant \PKzS, \PgL\ and \PagL\ 
samples we only weighted a small fraction of the total 
sample in order to reach a statistical accuracy better than the limits 
imposed by the systematic error.  
\section{Inverse slopes of the transverse mass spectra}
\begin{figure}[h]
{
\hbox to\hsize{\hss
\includegraphics[width=0.5\hsize]{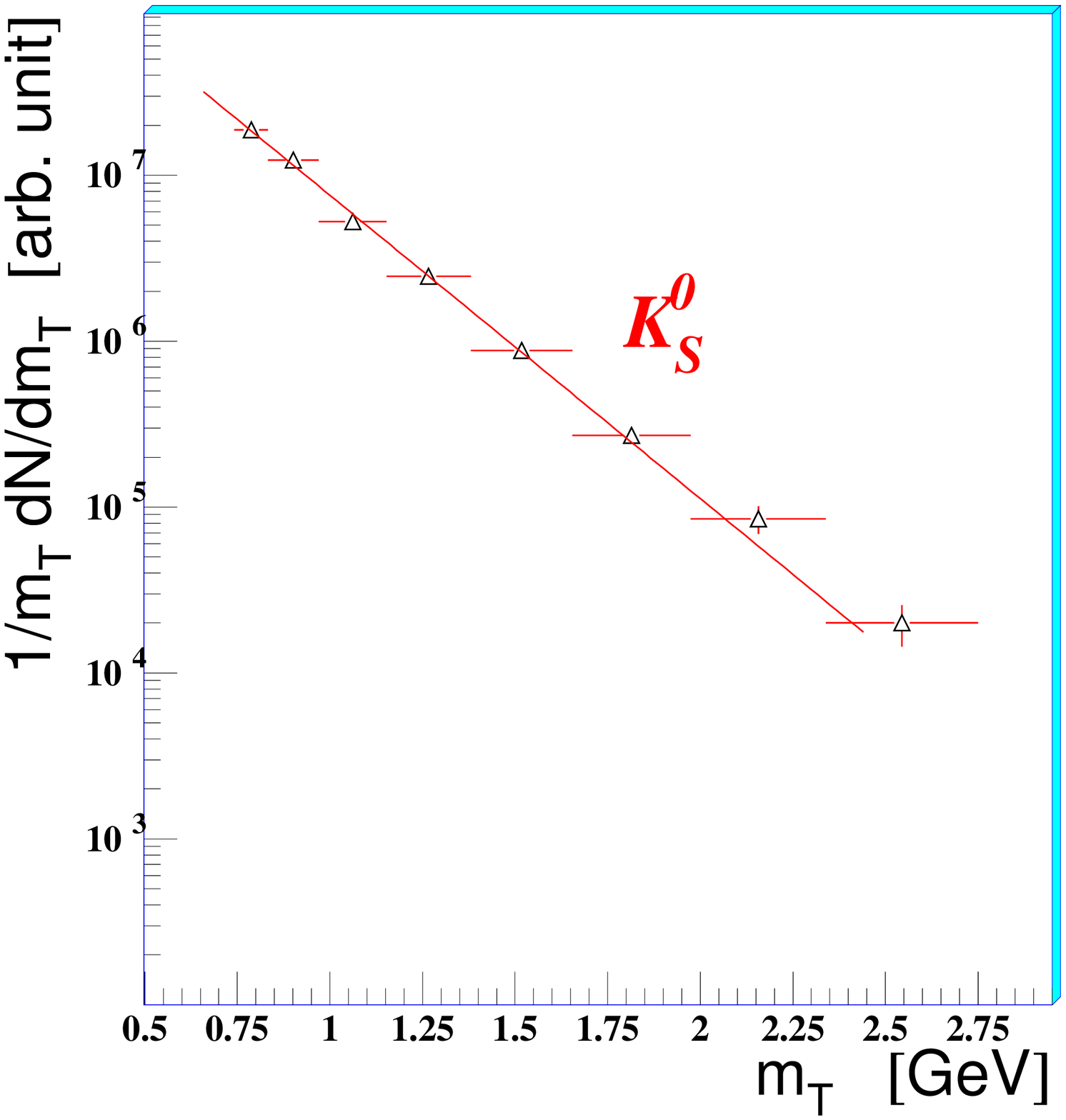} % quello finale (del paper)
\includegraphics[width=0.5\hsize]{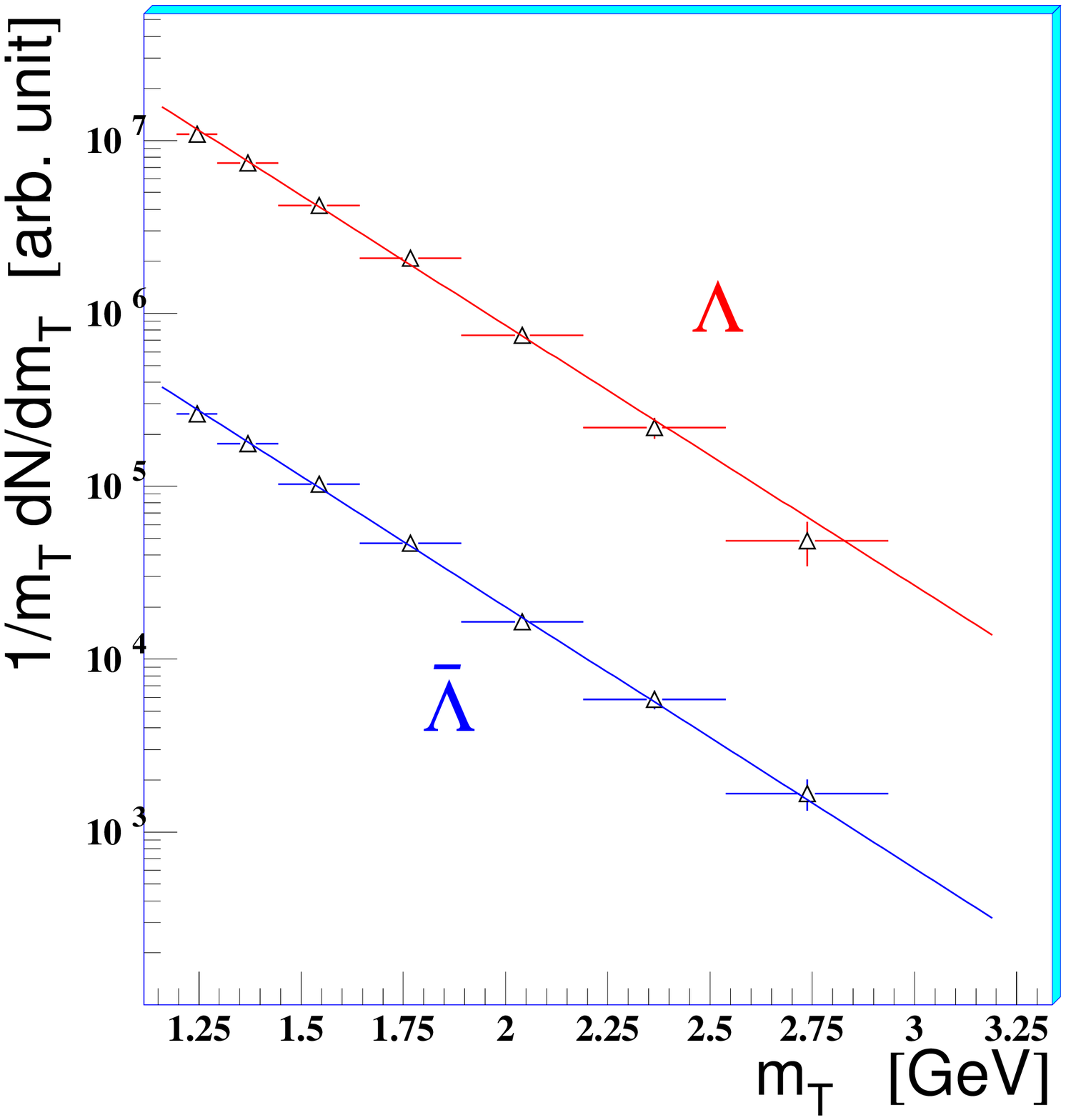}
\hss}} 
{
\hbox to\hsize{\hss
\includegraphics[width=0.5\hsize]{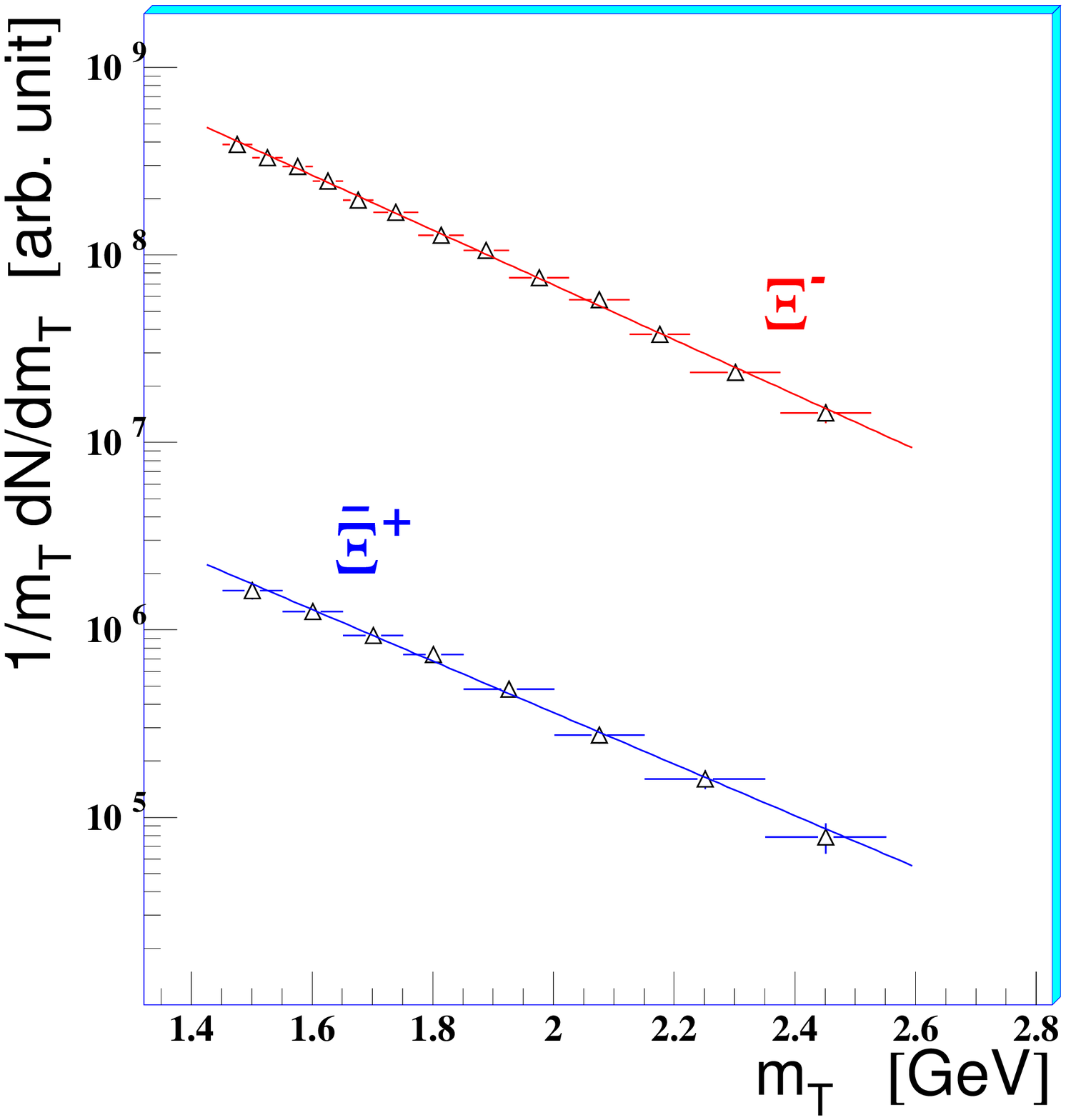}
\includegraphics[width=0.5\hsize]{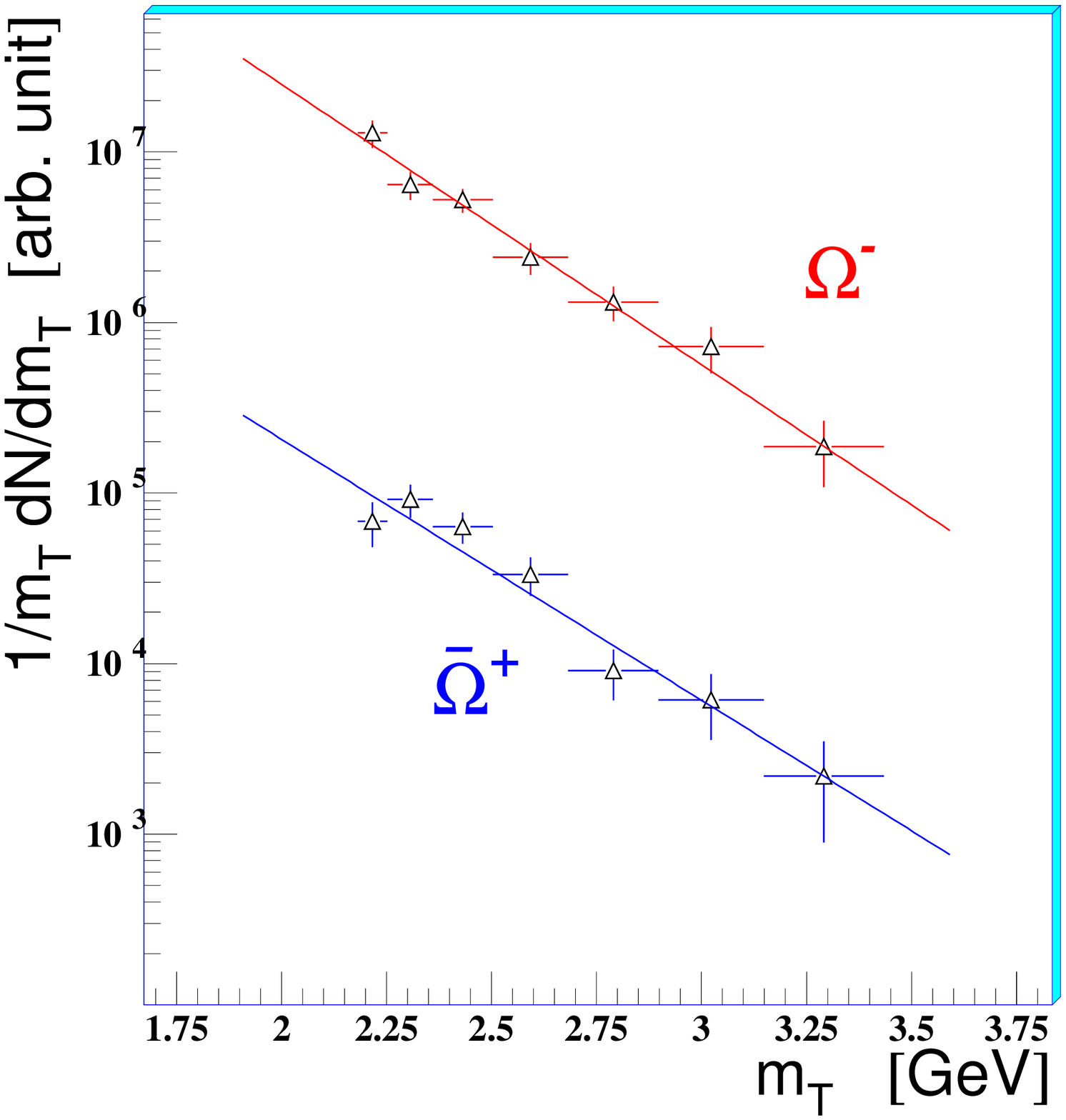} % quello finale (del paper)
\hss} }
\caption{Transverse mass spectra of strange particles 
 from about the 55\% most central Pb-Pb cross section at 158 $A$\ GeV/$c$.  
  The superimposed exponential functions have inverse slopes equal to the  
   $T_{app}$\ values obtained from the maximum likelihood fits.}  
\label{fig:all_spectra}
\end{figure}

The double-differential $(y,m_{\tt T})$\ distributions for each particle 
have been parametrized using the expression
\noindent
\begin{center}
\begin{equation}
\frac{d^2N}{m_{\tt T}\,dm_{\tt T} dy}=f(y) \hspace{1mm} \exp\left(-\frac{m_{\tt T}}{T_{app}}\right)
\label{eq:expo}
\end{equation}
\end{center}
assuming the rapidity distribution to be flat ($f(y)={\rm const}$)   
within our acceptance region.  
The inverse slope parameter $T_{app}$\ (``apparent temperature'') 
has been extracted by means of a maximum likelihood fit of Eq.~\ref{eq:expo} 
to the data. As discussed in detail in the next section, this apparent temperature 
can be interpreted as due to the  
thermal motion coupled with a collective transverse flow  
of the fireball components assumed to be in thermal equilibrium.  
\newline
Transverse mass distributions $1/m_{\tt T} \, dN/dm_{\tt T} $\ for strange particles
%from Pb-Pb collisions at 158 $A$\ GeV/$c$\
measured in the centrality
range accessible to the experiment are shown in figure~\ref{fig:all_spectra}.
The shapes of all spectra are well described by single exponential functions.
In the next section, we shall 
attempt a global fit with the blast-wave model.  
%exploit the small deviations from the exponential in the 
%attempt of disentangling the transverse flow from the thermal motion. 
\newline
The inverse slope parameters $T_{app}$\ of the transverse mass spectra are 
given in table~\ref{tab:InvSlopes}.  They are in agreement within the 
errors with those measured over a smaller centrality range 
(about 40\% of the Pb-Pb inelastic cross section) by the WA97 
experiment~\cite{MtWA97}. 
\begin{center}
\begin{table}[h]
\caption{Inverse slopes parameter $T_{app}$\ (MeV) of the strange particles
in the full centrality range ({\bf 0-4}). The first error is statistical, the
second one systematic. }%The \PgOm\ and \PagOp\ 
%inverse slopes are preliminary results.}
\label{tab:InvSlopes}
{\footnotesize
\begin{center}
\begin{tabular}{|c|c|}
\hline
{} &{}\\[-1.5ex]
{} & Inverse Slope (MeV)\\[1ex]
\hline
{} &{}\\[-1.5ex]
\PKzS & $237\pm4\pm24$  \\[1ex] % nuovi valori (quelli del paper)
%\PKzS & $229\pm9\pm24$  \\[1ex] % mostrato sul poster
\PgL  & $289\pm7\pm29$  \\[1ex]
\PagL & $287\pm6\pm29$  \\[1ex]
\PgXm & $297\pm5\pm30$  \\[1ex]
\PagXp& $316\pm11\pm30$ \\[1ex]
\PgOm & $264\pm19\pm27$ \\[1ex]  % nuovi valori (quelli del paper)
\PagOp& $284\pm28\pm27$ \\[1ex]  % nuovi valori (quelli del paper)
%\PgOm & $280\pm16\pm40$ \\[1ex]  % mostrato sul poster
%\PagOp& $324\pm29\pm40$ \\[1ex]  % mostrato sul poster
\hline
\end{tabular}
\end{center}
}
\vspace*{-13pt}
\end{table}
\end{center}
\section{Blast-wave description of the spectra}
\begin{figure}[hb]  
\hbox to\hsize{\hss
\includegraphics[width=0.85\hsize]{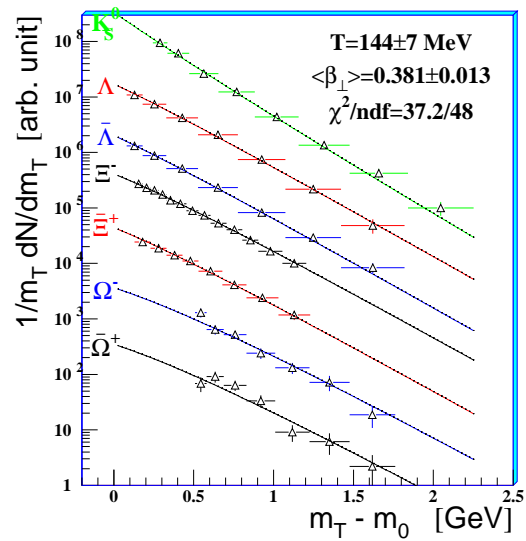} % that of the poster
\hss}
\caption{Transverse mass spectra fitted with collective flow model 
         for the linear ($n=1$) transverse velocity profile.  
         The quoted values are valid for the full NA57 centrality sample.}
\label{fig:blast}
\end{figure}

\begin{figure}[h]
\hbox to\hsize{\hss
\includegraphics[width=0.8\hsize]{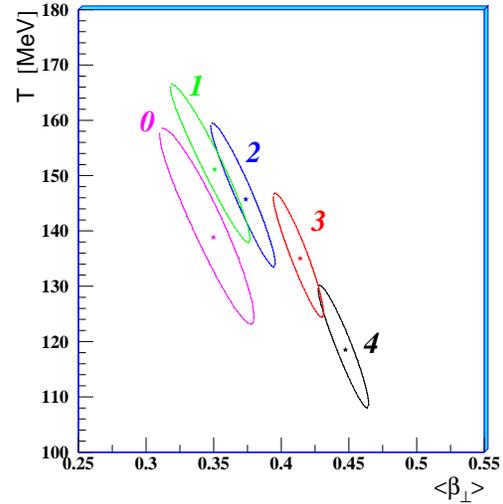} % that of the poster
\hss}
\caption{The 1$\sigma$\ confidence level contours from
         fits to individual centrality classes..}
\label{fig:cont_msd}
\end{figure}

We consider in this section the statistical hadronization model of
ref.~\cite{BlastRef} to describe the strange particle spectra
discussed above. 
The model assumes that particles decouple from a system in local thermal
equilibrium with temperature $T$, which expands both longitudinally
and in the transverse direction; the quantum statistical distributions are
approximated by the Boltzmann distribution.
The longitudinal expansion is boost-invariant, the transverse one 
is defined in terms of a transverse velocity field.  
\newline
The blast-wave model predicts a double differential cross section
for a given particle $j$\
%($j=\PKzS,\PgL, \PagL, \PgXm, \PagXp, \PgOm, \PagOp$)
of the form:
\noindent
\begin{center}
\begin{equation}
\frac{d^2N_j}{m_{\tt T} dm_{\tt T} dy} %\propto
    = \int_0^{R_G}{\mathcal{A}_j
     m_{\tt T} K_1\left( \frac{m_{\tt T} \cosh \rho}{T} \right)
     I_0\left( \frac{p_{\tt T} \sinh \rho}{T} \right) r dr} \;
\label{eq:blast}
\end{equation}
\end{center}
where $\rho(r)=\tanh^{-1} \beta_{\perp}(r)$, 
$K_1$\ and $I_0$\ are two modified Bessel functions and 
$R_G$\ is the transverse geometric (gaussian) radius of the source.  
\newline
%In this preliminary study, we consider a simplified model where 
%the transverse flow velocity is assumed to be constant all along its 
%profile: $\beta_{\perp}(r)=<\beta_{\perp}>$. A more detailed analysis 
%using different transverse velocity profiles and a study of the stability  
%of the fits (the parameters $T$\ and $<\beta_\perp>$\ are strongly 
%anti-correlated), as well as their sensitivity to individual particles 
%and the estimate of the systematic errors,  
%has been carried out after the time of this conference.  
%For the complete description of the transverse mass spectrum  
%analysis refere to the forthcoming publication~\cite{blastpaper}.  
In a preliminary study~\cite{Bruno2}, we considered a simplified model where 
the transverse flow velocity was assumed to be constant all along its 
profile: $\beta_{\perp}(r)=<\beta_{\perp}>$. A more detailed analysis 
using different transverse velocity profiles 
has now been performed.  
%has been carried out after the time of this conference. 
We have also studied the stability 
of the fits (the parameters $T$\ and $<\beta_\perp>$\ are strongly 
anti-correlated), as well as their sensitivity to individual particle spectra.  
%and estimated the systematic errors.  
A complete description of the transverse mass spectrum  
analysis at 160 $A$\ GeV/$c$\   
will be presented in a forthcoming publication~\cite{blastpaper}.   
%\newline
In this new study, the transverse velocity field $\beta_{\perp}(r)$\ 
has been parametrized according to power laws: 
\begin{equation}
%\[
\beta_{\perp}(r) = \beta_S \left[ \frac{r}{R_G} \right]^{n}
  \quad \quad \quad r \le R_G
  %\]
  \label{eq:profile}
  \end{equation}
%where $R_G$\ is the transverse geometric
%(gaussian) radius of the source and 
where 
the exponent $n$\ was assumed to be $n=0,1/2,1$~or~$2$.  
Assuming a uniform particle density, the average transverse 
velocity is calculated according to  
%\begin{equation}
$
<\beta_{\perp}> = \frac{2}{2+n}  \beta_S
\label{eq:averageB}
$
%\end{equation}
\newline
As shown in fig.~\ref{fig:blast}, 
the simultaneous best fit of eq.~\ref{eq:blast} 
%(the $n=1$\ profile only is considered in this paper)   
($n=1$\ profile case)  
to the data points of  all the measured strange particle spectra 
successfully describes the data with  
%all the distributions with %$\chi^2/ndf=26.3/39$, 
$\chi^2/ndf=37.2/48$,  
yielding the following values for the two 
basic quantities $T$\ and $<\beta_\perp>$: 
%\[
%T = 161 \pm 7 MeV \, , \quad \quad  <\beta_\perp>=0.395 \pm 0.015  \; .
%\nonumber \]
\begin{center}
\( T = 144 \pm 7{\rm (stat)}\pm14{\rm(syst)} \; {\rm MeV}  \) \newline
\( <\beta_\perp>= 0.381 \pm 0.013{\rm (stat)}\pm0.012{\rm(syst)}  \, . \)
\end{center}
The quadratic profile is disfavoured by the data,  
yielding the largest value of $\chi^2$. 
The other two profiles ($n=0$, $n=1/2$) yield  
similar values of the freeze-out temperatures and of the  average transverse 
flow velocities, as the $n=1$\ case,   
with good values of $\chi^2/ndf$.  For the study of the centrality dependence  
of the freeze-out parameters presented below, the linear profile has been used.  
\newline
The particles have been divided into two groups --- those which share
quarks in common with the nucleons and those which do 
not~\footnote{It is known that the particles of the two groups may 
exhibit different production features, e.g. in the rapidity spectra.}  --- 
and the fit procedure has been repeated separately for the two groups. 
The results of such fits are summarized in table~\ref{tab:Blast1}. 
They suggest similar freeze-out conditions for the two groups. %of particles.  
Since the interaction cross-sections for 
the particles of the two groups are quite different, this finding would suggest 
a similar production mechanism %for the particles of the two groups
and final state interactions of limited importance 
(e.g. a rapid thermal freeze-out). 
A similar conclusion concerning the evolution of the system was reached in the 
study of the HBT correlation functions of negative pions at high 
$p_{\tt T}$\ in the WA97 
experiment~\cite{HBTpaper}.
\begin{center}
\begin{table}[h]
\caption{Thermal freeze-out temperature $T$\ 
and average transverse flow velocity
$<\beta_\perp>$\ in the full centrality range. 
The first error is statistical, the second one systematic.}
\label{tab:Blast1}
{\footnotesize
\begin{center}
\begin{tabular}{|c|c|c|c|}
\hline
{} &{} &{} &{}\\[-1.5ex]
{} & $T$\ (MeV) & $<\beta_\perp>$ & $\chi^2/ndf$\\[1ex]
\hline
{} &{} &{} &{}\\[-1.5ex]
%$\PKzS,\PgL,\PgXm$          & $160\pm10$ & $0.39\pm0.02$ & $12.7/15$ \\[1ex] % poster
%$\PagL,\PagXp,\PgOm,\PagOp$ & $167\pm16$ & $0.39\pm0.03$ & $12.2/22$ \\[1ex] % poster
%$\PKzS,\PgL,\PgXm$          & $146\pm8\pm14$  & $0.376\pm0.015\pm0.012$ & $18.1/23$ \\[1ex] % paper
%$\PagL,\PagXp,\PgOm,\PagOp$ & $130\pm28\pm14$ & $0.403\pm0.032\pm0.012$ & $18.5/23$ \\[1ex] % paper
$\PKzS,\PgL,\PgXm$          & $146\pm8\pm14$  & $0.38\pm0.01\pm0.01$ & $18.1/23$ \\[1ex] % paper
$\PagL,\PagXp,\PgOm,\PagOp$ & $130\pm28\pm14$ & $0.40\pm0.03\pm0.01$ & $18.5/23$ \\[1ex] % paper
\hline
\end{tabular}
\end{center}
}
\vspace*{-13pt}
\end{table}
\end{center}
In order to see whether some species deviate from
the common temperature and transverse flow velocity scenario
determined above, we have attempted to fit the spectra 
separately. For statistical reasons, \PgOm\ and \PagOp\ 
spectra have been combined. For all particle species,  
$T$\ and $<\beta_{\perp}>$\ do not deviate more than two standard  
deviations from the common freeze-out scenario.  
\newline
We have also performed the global fit to the spectra 
%for different centrality classes: 
%in this preliminary analysis 
%we consider three centralities only, in order to reduce the 
%statistical error, namely our (5\%) most central collision class $4$, 
%classes $0$\ and $1$\ merged (most peripheral group) and 
%classes $2$\ and $2$\ merged (intermediate centrality).  
for each of the five NA57 centrality classes. 
The results are summarized in table~\ref{tab:blastmsd}.  
\begin{center}
\begin{table}[h]
\caption{Thermal freeze-out temperature ($T$) and average 
transverse flow velocity ($<\beta_\perp>$) in five centrality classes,  
assuming a linear transverse profile.}  
\label{tab:blastmsd}
{\footnotesize
\begin{center}
\begin{tabular}{|c|c|c|c|}
\hline
{} &{} &{} &{}\\[-1.5ex]
{} & $T$\ (MeV) & $<\beta_\perp>$ & $\chi^2/ndf$\\[1ex]
\hline
{} &{} &{} &{}\\[-1.5ex]
%$0-1$ & $163\pm16$ & $0.35\pm0.03$ & $37.8/37$ \\[1ex] % poster  
%$2-3$ & $167\pm13$ & $0.40\pm0.03$ & $32.6/37$ \\[1ex] % poster
%$ 4 $ & $131\pm10$ & $0.47\pm0.02$ & $37.4/37$ \\[1ex] % poster
$0$ & $139\pm25$ & $0.35\pm0.04$ & $33.2/35$ \\[1ex] % paper
$1$ & $151\pm15$ & $0.35\pm0.03$ & $42.3/42$ \\[1ex] % paper
$2$ & $146\pm17$ & $0.37\pm0.03$ & $41.3/43$ \\[1ex] % paper
$3$ & $135\pm14$ & $0.41\pm0.02$ & $31.8/41$ \\[1ex] % paper
$4$ & $118\pm13$ & $0.45\pm0.02$ & $52.7/43$ \\[1ex] % paper
\hline
\end{tabular}
\end{center}
}
\vspace*{-13pt}
\end{table}
\end{center}
In figure~\ref{fig:cont_msd} we show the $1\sigma$\ confidence level contours
in the $<\beta_\perp>$--$T$\ parameter space as
obtained from the blast-wave fit with the $n=1$\  profile case.  
Apart from the most peripheral interactions, 
the centrality dependence  of the two parameters shows opposite 
trends, the temperature is decreasing and the average flow velocity 
is increasing with increasing centrality.  
\newline
\section{Conclusions and outlook}
We have analysed the transverse mass spectra of
high statistics, high purity samples of \PKzS,  \PgL, $\Xi$\ and $\Omega$\
particles produced in Pb-Pb collisions at 158 $A$\ GeV/$c$\ over a wide range
of collision centrality (about the most central 55\% of the Pb-Pb interaction
cross section). \newline
For each hyperon species, the $m_{\tt T}$\ slope is found to be in good agreement
with that of the corresponding anti-hyperon. \newline  
The analysis %of the transverse mass spectra at 158 $A$\ GeV/$c$\
in the framework of a blast-wave  
model using a linear velocity profile suggests  that after a central 
collision the system expands explosively; 
the system then freezes-out when the temperature is of the order
of 120 MeV with an average transverse  flow 
velocity of about one half of the speed of light.  
Particles with and without valence quarks in common with the nucleon
appear to have similar behaviour.   
With increasing collision centrality,  
the transverse flow velocity increases and the final temperature decreases.

\end{document}